\documentclass[prl,aps,superscriptaddress,showpacs,twocolumn,amssymb,amsmath,amstext,amsfont]{revtex4}
\usepackage{graphicx}
\usepackage{theorem}
\usepackage{dcolumn}
\usepackage{bm}
\usepackage{mathrsfs}
\usepackage{setspace}

\begin{document}

\title{Determination of the effective kinematic viscosity for the decay of quasiclassical turbulence in superfluid $^4$He}
\affiliation{National High Magnetic Field Laboratory, 1800 East Paul Dirac Drive, Tallahassee, FL 32310, USA}
\affiliation{Mechanical Engineering Department, Florida State University, Tallahassee, FL 32310, USA}
\affiliation{School of Physics and Astronomy, University of Birmingham, Birmingham B15 2TT, United Kingdom}

\author{J. Gao}
\affiliation{National High Magnetic Field Laboratory, 1800 East Paul Dirac Drive, Tallahassee, FL 32310, USA}
\affiliation{Mechanical Engineering Department, Florida State University, Tallahassee, FL 32310, USA}

\author{W. Guo \footnote{Corresponding: wguo@magnet.fsu.edu}}
\affiliation{National High Magnetic Field Laboratory, 1800 East Paul Dirac Drive, Tallahassee, FL 32310, USA}
\affiliation{Mechanical Engineering Department, Florida State University, Tallahassee, FL 32310, USA}

\author{W.F. Vinen}
\affiliation{School of Physics and Astronomy, University of Birmingham, Birmingham B15 2TT, United Kingdom}

\date{\today}

\begin{abstract}
The energy dissipation of quasiclassical homogeneous turbulence in superfluid $^4$He (He II) is controlled by an effective kinematic viscosity $\nu'$, which relates the energy decay rate $dE/dt$ to the density of quantized vortex lines $L$ as $dE/dt=-{\nu'}({\kappa}L)^2$. The precise value of $\nu'$ is of fundamental importance in developing our understanding of the dissipation mechanism in He II, and it is also needed in many high Reynolds number turbulence experiments and model testing that use He II as the working fluid. However, a reliable determination of $\nu'$ requires the measurements of both $E(t)$ and $L(t)$, which was never achieved. Here we discuss our study of the quasiclassical turbulence that emerges in the decay of thermal counterflow in He II at above 1 K. We were able to measure $E(t)$ using a recently developed flow visualization technique and $L(t)$ via second sound attenuation. We report the $\nu'$ values in a wide temperature range determined for the first time from a comparison of the time evolution of $E(t)$ and $L(t)$.
\end{abstract}

\pacs{67.25.dk, 29.40.Gx, 47.27.-i} \maketitle

Below about 2.17 K, liquid $^4$He transits to the superfluid phase (He II) in which an inviscid irrotational superfluid component (i.e. the condensate) coexists with a viscous normal-fluid component (i.e. the thermal excitations) \cite{Tilley-1986-book}. The fraction of the normal fluid drops drastically with decreasing temperature and only amounts to about 0.7\% of the total density at 1 K \cite{Donnelly-1998}. This quantum fluid system exhibits fascinating hydrodynamic properties. For instance, the rotational motion of the superfluid in a simply-connected volume can occur with the formation of topological defects in the form of vortex lines. These vortex lines all have identical cores with a radius $a_0\simeq$1 {\AA} and they each carry a single quantum of circulation $\kappa$=10$^{-3}$ cm$^2$/s \cite{Donnelly-1991-book}. Turbulence in the superfluid therefore takes the form of an irregular tangle of vortex lines (quantum turbulence). Turbulence in the normal fluid is expected to be more similar to that in a classical fluid, but a force of mutual friction between the two fluids, arising from the scattering of thermal excitations by the vortex lines, can affect the flows in both fluids \cite{Vinen-2002-JLTP}.

At above 1 K, despite being a two-fluid system with many properties restricted by quantum effects, He II is observed to behave very similarly to classical fluids when a turbulent flow is generated by methods conventionally used in classical fluid research, such as by a towed grid \cite{Stalp-1999-PRL} or a rotating propeller \cite{Maurer-1998-EPL}. Even in a non-classical thermal counterflow induced by an applied heat current in He II, it has been revealed that quasiclassical turbulence can emerge during the decay of counterflow after the heat current is switched off \cite{Skrbek-2003-PRE,Chagovets-2007-PRE,Gao-2015-JETP}. The quasiclassical behavior of He II is interpreted as the consequence of a strong coupling of the two fluids by mutual friction at large scales \cite{Vinen-2000-PRB}. It is suggested that the turbulent eddies in the normal fluid are matched by eddies in the superfluid produced by polarized vortices \cite{Barenghi-1997-PhysFlu,Baggaley-2012-EPL}, although different views regarding the bundling of the vortices exist \cite{Volovik-2003-JETP,Tsubota-2004-PRB,Kivotides-2011-JFM}. The coupled fluids behave at large length scales like a single-component viscous fluid at high Reynolds number. At small scales, due to the viscous dissipation in the normal fluid and the discrete vortex-line structure in the superfluid, the flows in the two fluids become decoupled. Mutual friction dissipation sets in at these small scales. This mechanism of coupling becomes weaker at lower temperatures. Nevertheless, at temperatures below 0.5 K where the normal-fluid fraction is essentially zero, quasiclassical turbulence in the superfluid was still observed \cite{Walmsley-2014-PNAS, Zmeev-2015-PRL}. In this case, it is generally believed that a classical Richardson cascade of the turbulence energy in the superfluid exists at scales greater than the mean intervortex distance $\ell$=$L^{-1/2}$ (where $L$ is the line density, i.e. vortex length per unit volume). But unlike at higher temperatures, this energy cascade can no longer be terminated by mutual friction dissipation. Instead, the turbulence energy is further transferred down to smaller scales via a cascade of Kelvin wave excitations on the vortices, which eventually leads to phonon emission \cite{Vinen-2001-PRB,Vinen-2003-PRL}.

The classical behavior of He II, especially in the two-fluid regime at above 1 K, has brought up the feasibility of using He II in classical turbulence research and for practical model testing. He II has very small kinematic viscosity which allows the generation of flows with extremely high Reynolds numbers that can hardly be achieved with other conventional fluid materials \cite{Skrbek-1999-JP}. Various projects have been launched for this purpose \cite{Donnelly-1991-Collection, Fuzier-2001, Michel-2014}. However, the viscosity that controls the energy dissipation of the quasiclassical turbulence above 1 K is not the normal-fluid viscosity but instead an effective kinematic viscosity $\nu'$ that accounts for both the viscous dissipation in the normal fluid and the mutual friction dissipation at small scales. The precise value of $\nu'$ is needed in the design of these He II based quasiclassical turbulence experiments. Furthermore, making reliable measurements of the $\nu'$ values will be indispensable in rigorously testing the various theories about the dissipation mechanism in He II \cite{Vinen-2002-JLTP, Stalp-1999-PRL, Skrbek-2012-PhysFlu}, which will be fundamentally important in advancing our knowledge of quantum turbulence.

Stalp \emph{et al.} first introduced $\nu'$ in a theoretical model for interpreting their measured vortex density $L(t)$ during the decay of a towed-grid generated turbulence in He II above 1 K \cite{Stalp-1999-PRL}. By analogy with the energy decay equation for classical turbulence \cite{Henze-1975-book}, Stalp \emph{et al.} proposed that the total turbulence kinetic energy per unit He II mass, $E(t)$, decays as $dE/dt=-{\nu'}({\kappa}L)^2$ \cite{Stalp-1999-PRL}. Approximately, $E$ can be evaluated as $E=E_1+E_2$, where $E_1$ comes from the flows in the superfluid on scales at or below $\ell$ associated with individual vortex lines and $E_2$ represents the kinetic energy density associated with large-scale flows in the coupled turbulence. $E_1$ can be estimated as $E_1$=$B(\rho_s/\rho){\kappa}^{2}L$ \cite{Donnelly-1991-book}, where the dimensionless factor $B\simeq{\frac{1}{4\pi}}$\texttt{ln}$(l/a_0)$ is typically about unity, and $\rho_s/\rho$ denotes the ratio of the superfluid density to the total density of He II. $E_2$ can be evaluated as $E_2=\frac{1}{2}(\Delta{U})^2$, where $\Delta{U}=\langle(U-\overline{U})^2\rangle^{1/2}$ denotes the root mean square velocity fluctuation of the coupled flows. For quasiclassical turbulence in He II, $E_2$ is normally much greater than $E_1$ \cite{Vinen-2000-PRB}. The energy decay rate equation can therefore be formally written as
\begin{equation}
\frac{dE}{dt}=B{\kappa}^{2}\frac{\rho_s}{\rho}\frac{dL}{dt}+\frac{dE_2}{dt}=-{\nu'}({\kappa}L)^2
\label{eqn1}
\end{equation}
Based on Eq.~(\ref{eqn1}), Stalp \emph{et al.} neglected the $E_1$ contribution and derived an explicit expression for $L(t)$ at large decay times as
\begin{equation}
L(t)\simeq\frac{D(3C)^{3/2}}{2\pi\kappa\sqrt{\nu'}}{\cdot}t^{-3/2}
\label{eqn-L}
\end{equation}
where $D$ is the width of the flow channel and $C=1.5$ is the Kolmogorov constant \cite{Stalp-1999-PRL}. The derivation of Eq.~(\ref{eqn-L}) involves two major hypotheses: 1) the size of the energy-containing eddy is saturated by the channel width $D$; and 2) the coupled turbulence has a classical Kolmogorov energy spectrum that extends to \emph{all} scales (i.e. $E_2$=$\int\widetilde{E}_2(k)dk$, where $\widetilde{E}_2(k)$ depends on the wave number $k$ as $\widetilde{E}_2(k)\propto{k^{-5/3}}$) \cite{Stalp-1999-PRL}. These hypotheses are, to some extend, supported by the observed $L(t)\propto{t}^{-3/2}$ behavior at large decay times. The value of $\nu'$ was then determined by fitting the measured $L(t)$ using Eq.~(\ref{eqn-L}) \cite{Stalp-2002, Niemela-2000-JLTP}. This method was later used by other groups for estimating $\nu'$ in decaying counterflow and decaying co-flow in a channel \cite{Chagovets-2007-PRE, Babuin-2015-PRB}. A similar idea was also applied to the study of quasiclassical turbulence in pure superfluid at very low temperatures where the effective viscosity $\nu'$ arises from completely different dissipation mechanism \cite{Walmsley-2007-PRL, Walmsley-2008-PRL, Bradley-2008-PRL}. Nevertheless, in all these studies the $\nu'$ values obtained using Eq.~(\ref{eqn-L}) are indeed dubious as discussed by Zmeev \emph{et al.} \cite{Zmeev-2015-suppl}. There is no evidence showing that the energy-containing eddy size must be the same as the channel width $D$. Although intuitively they should not be too different, any possible difference can result in significant change in the fitted $\nu'$ values since $\nu'\propto{D^2}$ according to Eq.~(\ref{eqn-L}). Furthermore, if there is indeed a Kolmogorov spectrum, this spectrum must break down near the cut-off scale $D$. We also note that Skrbek's group managed to evaluate $\nu'$ in steady-state co-flows \cite{Babuin-2014-EPL}. But their analysis requires additional assumptions which further limit the result accuracy. A reliable determination of $\nu'$ for quasiclassical turbulence in He II can be made only if one can measure directly both $L(t)$ and $E_2(t)$.

In He II, the vortex density $L(t)$ can be readily measured using either second sound attenuation \cite{Stalp-1999-PRL,Vinen-1957-PRS} or trapping of negative ions \cite{Walmsley-2007-PRL, Awschalom-1984-PRL}. However, a direct measurement of the turbulence energy is challenging. Typical measurement tools for $\Delta{U}$, such as pitot pressure tubes \cite{Maurer-1998-EPL}, normally have limited spatial resolution, and their application requires a large mean flow velocity which is not always present in decaying turbulence. Another route to probe the turbulence energy is to measure the resulting heat input to the fluid as the turbulence decays \cite{Samuels-1998-PRL}. Along this line, Bradley \emph{et al} developed a unique Andreev scattering technique and made the first direct measurement of the energy decay in superfluid $^3$He-B at zero temperature limit \cite{Bradley-2011-Nat}. However, they could not determine the vortex density in the same experiment and thus cannot deduce the values for $\nu'$. In this paper, we report the measurements of both $L(t)$ and $E_2(t)$ in decaying counterflow turbulence in He II, by combining the second sound attenuation technique and a recently developed tracer-line tracking flow visualization technique \cite{Gao-2015-RSI,Marakov-2015-PRB}. Our method is applicable to the two-fluid regime at above 1 K. We show that a reliable determination of the $\nu'$ values can be made.

\begin{figure}[htb]
\includegraphics[scale=0.4]{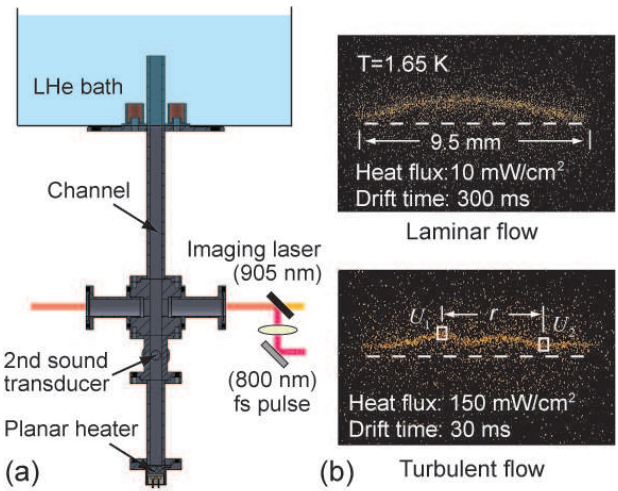}
\caption{(color online). (a) Schematic diagram of the experimental setup. (b) Typical images showing the deformation of the He$_{2}^{*}$ molecular tracer lines in steady-state thermal counterflow. The white dashed lines indicate the initial locations of the tracer lines.} \label{Fig-1}
\end{figure}

Our experimental setup is shown in Fig.~\ref{Fig-1}~(a). A stainless steel channel (square cross-section: (9.5~mm)$^{2}$; length 300~mm) is attached to a pumped helium bath whose temperature can be controlled within 0.1 mK. A planar heater at the lower end of the channel can be used to drive a thermal counterflow, i.e. the superfluid flowing towards the heater and the normal fluid away from it \cite{Landau-book}. The mean velocity $\overline{U}$ of the normal fluid is related to the heat flux $q$ by $\overline{U}=q/\rho{s}T$, where $s$ is the specific entropy of the helium. When the heat flux is greater than a small critical value, it is known that the superfluid can become turbulent and a self-sustained vortex tangle is generated by the mutual friction between the two fluids \cite{Vinen-2002-JLTP}. We have reported that the normal fluid can also become turbulent above a threshold heat flux $q_c$ (e.g. $q_c\sim60$ mW/cm$^2$ for 1.65 K), exhibiting a novel $k^{-2}$ energy spectrum \cite{Marakov-2015-PRB}. This energy spectrum is likely caused by the mutual friction dissipation that occurs in a wide range of scales in the normal fluid, since the two fluids have opposite mean velocities and therefore cannot get completely coupled. As the heater is turned off, the heat current decays to zero with a thermal time constant $\tau$ \cite{Vinen-1957-PRS,Gordeev-2005-JLTP}. In the absence of the heat current, the two fluids can then get coupled at large scales by mutual friction. The time it takes to establish the coupling can be estimated using the formula derived by Vinen \cite{Vinen-2000-PRB} and is typically in the range of 1-10 ms in our experiment. Our analysis on $\nu'$ will relate to times that are greater than both the thermal time constant and the time required for complete coupling.

In order to extract quantitative flow field information, we have adopted a recently developed flow visualization technique by tracking thin lines of He$_{2}^{*}$ molecular tracers. These tracers are created via ionizing ground state helium atoms using a focused femtosecond laser pulse \cite{Gao-2015-RSI}. Above 1 K, He$_{2}^{*}$ tracers are completely entrained by the normal fluid and can be imaged via laser-induced fluorescence \cite{Rellergert-2008-PRL,Guo-2009-PRL,Guo-2010-PRL,Guo-2014-PNAS}. Fig.~\ref{Fig-1}~(b) shows typical images of the He$_{2}^{*}$ tracer lines in steady-state counterflow at 1.65 K. The streamwise velocity field of the normal fluid can be determined from the vertical displacements of the line segments \cite{Gao-2015-RSI}. Using this tracer-line tracking technique, we can probe the normal fluid motion (and hence the coupled-fluid motion) at scales from the channel width ($\sim$ 1 cm) down to about half the thickness of the tracer line ($\sim$ 100~$\mu$m).

\begin{figure}[htb]
\includegraphics[scale=0.65]{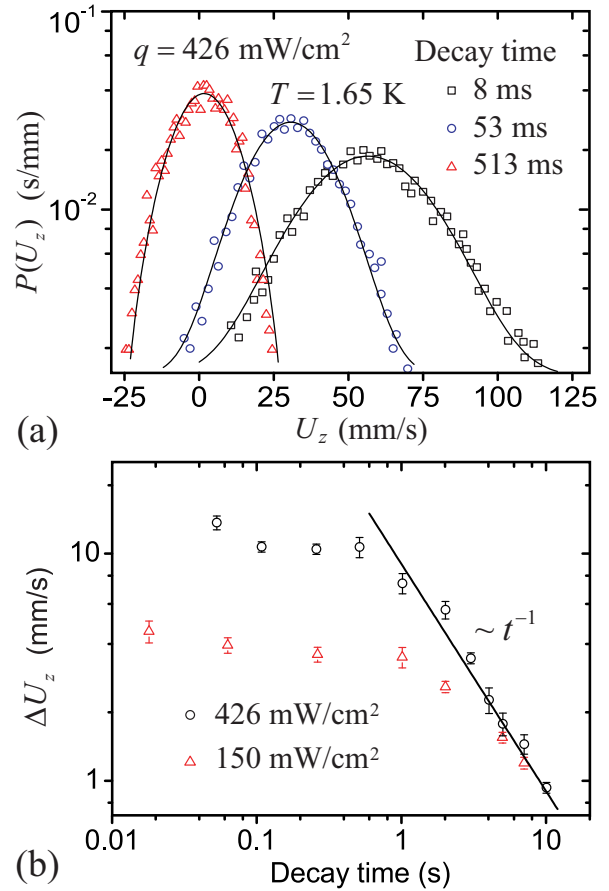}
\caption{(color online). (a) Velocity probability density functions (PDFs) in decaying counterflow turbulence at 1.65 K with an initial heat flux of 426 mW/cm$^2$. The solid curves represent Gaussian fits to the data. (b) Streamwise velocity fluctuation $\Delta{U_z}$ determined from the Gaussian fits of the velocity PDFs.}\label{Fig-2}
\end{figure}
For a given heat flux $q$, we normally maintain a steady-state counterflow for over 20 s and then switch off the heat current. We repeat the experiment 200 times and analyze the 200 images acquired at every decay time to produce velocity probability density functions (PDFs). Typical results for an initial heat flux of $q$=426 mW/cm$^2$ at 1.65 K are shown in Fig.~\ref{Fig-2}~(a). These velocity PDFs can be well fitted with Gaussian functions, which allow us to determine the time evolution of both the mean flow velocity $\overline{U}$ and the streamwise root mean square velocity fluctuation $\Delta{U_z}$. The time taken for $\overline{U}$ to decay to nearly zero is about 100 ms for $q$=426 mW/cm$^2$ and shorter at lower heat fluxes, in agreement with the expected thermal time constant. The measured decay of $\Delta{U_z}$ for typical initial heat fluxes is shown in Fig.~\ref{Fig-2}~(b). We observe that after the two fluids get coupled, the decay of $\Delta{U_z}$ is very slow and nearly flattens off at relatively small decay times. At large decay times, $\Delta{U_z}\propto{t^{-1}}$ and hence the energy $E_2(t)\propto{t^{-2}}$. The late decay behavior is in accordance with the decay of a quasiclassical turbulence with a Kolmogorov spectrum \cite{Stalp-1999-PRL}, but the initial flattening is more severe than the expected $(t+t_0)^{-1}$ behavior.

\begin{figure}[htb]
\includegraphics[scale=0.46]{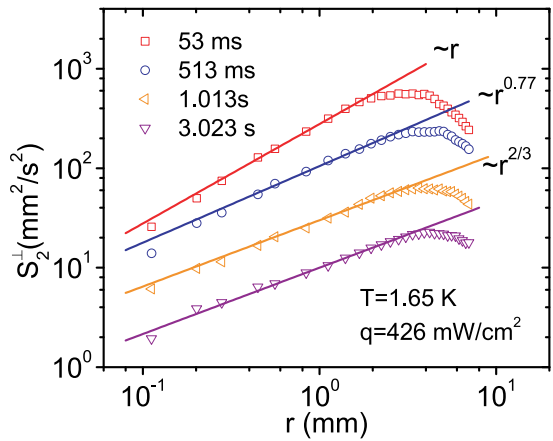}
\caption{(color online). The calculated 2nd order transverse structure function at different decay times in decaying counterflow with an initial heat flux of 426 mW/cm$^2$.}\label{Fig-3}
\end{figure}
We note in passing that the energy spectrum of the coupled turbulence can be directly probed in our experiment by calculating the second-order transverse structure function $S^{\perp}_2(r)=\langle(U_1-U_2)^2\rangle$ \cite{Marakov-2015-PRB}, where $r$ is the separation of two line segments (see Fig.~\ref{Fig-1}~(b)). The time evolution of the calculated $S^{\perp}_2(r)$ is shown in Fig.~\ref{Fig-3}. We observe that $S^{\perp}_2(r)\propto{r^n}$ below a few millimeters. This exponent $n$ leads to an energy spectrum $\widetilde{E}_2(k)\sim{k^{-(n+1)}}$ \cite{Kolmogorov-1941,Frisch-1995-book}. The observed variation of $n$ reveals that the coupled turbulence evolves from a non-classical form at small decay times with a spectrum close to that in steady state (i.e. $n$=1) to a quasiclassical turbulence at large decay times with a Kolmogorov spectrum (i.e. $n$=2/3). This spectrum transition is found to be responsible for the initial slow decay of $\Delta{U_z}$ and $E_2(t)$ \cite{Gao-2015-JETP}.

\begin{figure}[htb]
\includegraphics[scale=0.6]{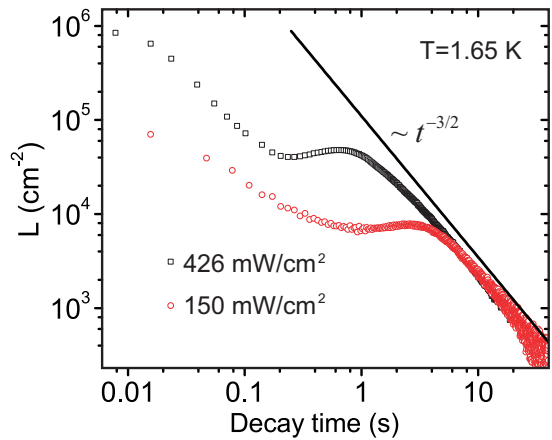}
\caption{(color online). The decay of the vortex line density $L(t)$ measured at different initial heat fluxes at 1.65K} \label{Fig-4}
\end{figure}
We also measured the vortex-line density $L(t)$ in decaying counterflow using the standard second sound attenuation method \cite{Vinen-1957-PRS}. The typical decay behavior of $L(t)$ at 1.65 K is shown Fig.~\ref{Fig-4}. We observe that when the normal fluid is turbulent in the steady state, the decay of $L(t)$ always exhibits three distinct regimes. The first regime occurs at very short decay times where $L(t)$ decays fast and in accordance with Vinen's phenomenological model \cite{Vinen-1957-PRS}. Subsequently, $L(t)$ can grow with time and show a ``bump" structure. At large decay times, $L(t)\propto{t^{-3/2}}$. This $L(t)$ decay behavior was reported in the past \cite{Vinen-1957-PRS, Skrbek-2003-PRE}. Skrbek \emph{et al} first realized that the $t^{-3/2}$ behavior at large decay times indicated the decay of a quasiclassical turbulence in the coupled two fluids, similar to those generated by a towed grid \cite{Skrbek-2003-PRE}. However, the underlying mechanism for the appearance of the bump and the switching to the $t^{-3/2}$ decay was unclear for many years despite various theoretical efforts \cite{Schwarz-1991-PRL,Nemirovskii-1994-Cryo,Sciacca-2010-PRB}. With the aid of our flow visualization, we have recently elucidated that the energy spectrum transition in the coupled turbulence is responsible for the observed complex $L(t)$ behavior \cite{Gao-2015-JETP}.

\begin{figure}[htb]
\includegraphics[scale=0.5]{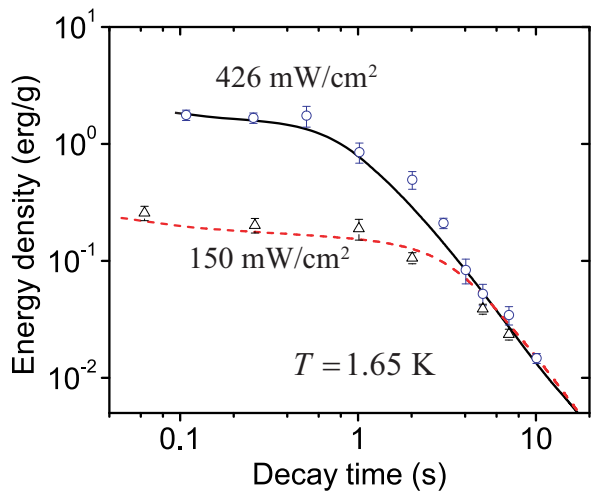}
\caption{(color online). The decay of the total turbulence energy density in decaying counterflow turbulence. The blue circles and black triangles are calculated based on the expression on the left-hand side of Eq.~(\ref{eqn2}). The black solid curve and the red dashed curve represent the results calculated using the integral on the right-hand side of Eq.~(\ref{eqn2}). The best agreement of the calculated energy density at large decay times is achieved with $\nu'/\kappa=0.46$.} \label{Fig-5}
\end{figure}

In order to determine the effective kinematic viscosity $\nu'$, we integrate Eq.~(\ref{eqn1}) from $t$ to infinity on both sides and write the total energy density $E(t)$ as
\begin{equation}
B\kappa^{2}\frac{\rho_s}{\rho}L(t)+E_2(t)={\nu'}\cdot\int^{\infty}_{t}{\kappa}^2L^2(t')dt'
\label{eqn2}
\end{equation}
Here $E_2(t)$ can be evaluated as $E_2(t)=\frac{3}{2}(\Delta{U_z})^2$, assuming that the large-scale turbulence in decaying counterflow is isotropic. This assumption should hold reasonably well at least at large decay times where the coupled flow shows a Kolmogorov spectrum for isotropic turbulence. The total turbulence energy density $E$ can be calculated based on our measured $L(t)$ and $\Delta{U_z}$ using the expression on left-hand side of Eq.~(\ref{eqn2}). The results for $q$=426 mW/cm$^2$ and $150$ mW/cm$^2$ at 1.65 K are shown in Fig.~\ref{Fig-5} as circles and triangles, respectively. For both heat fluxes, the contribution from $E_2$ dominates. The solid curve and the dashed curve shown in Fig.~\ref{Fig-5} are calculated based on the integral on the right-hand side of Eq.~(\ref{eqn2}). To evaluate this integral, we assume that the $t^{-3/2}$ behavior of $L(t)$ continues for decay times beyond the maximum measurement time in our experiment (about 40 s). Due to the fast decay of $L(t)$, the contribution to the integral at very long decay times is negligible. We then vary $\nu'$ and determine its value by requiring that the energy densities calculated with the expressions on either side of Eq.~(\ref{eqn2}) give the best agreement at large decay times. At 1.65 K, $\nu'/\kappa=0.46$ is obtained. We note that the calculated energy density curves indeed also show good agreement at relatively small decay times when the energy spectrum of the coupled turbulence still undergoes the transition.

\begin{figure}[htb]
\includegraphics[scale=0.55]{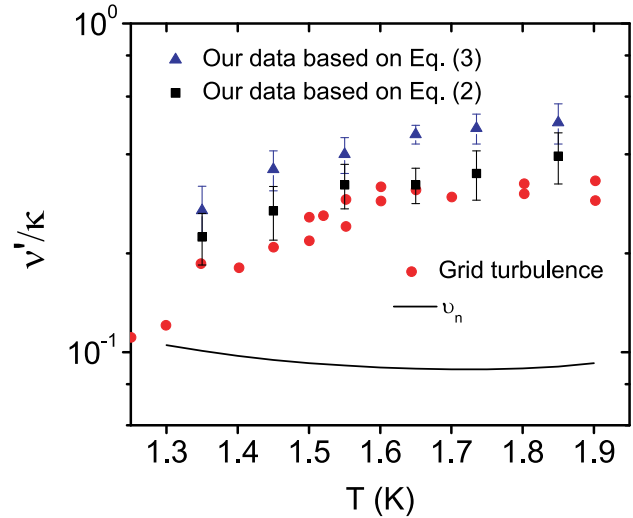}
\caption{(color online). Effective kinematic viscosity in units of $\kappa$. The blue triangles represent $\nu'$ values calculated using our new method via Eq.~(\ref{eqn2}). The black squares are calculated using our vortex density data via Eq.~(\ref{eqn-L}). The red solid circles represent $\nu'$ values obtained by Stalp \emph{et al.} in the towed-grid turbulence experiment \cite{Stalp-2002} that are corrected by Chagovets \emph{et al.} \cite{Chagovets-2007-PRE}. The black solid curve is the kinematic viscosity of He II calculated based on the normal-fluid viscosity alone \cite{Donnelly-1998}.} \label{Fig-6}
\end{figure}

We have made similar measurements in decaying counterflow at other temperatures above 1 K. The overall decay behaviors of the vortex density $L(t)$ and the root mean square velocity fluctuation $\Delta{U_z}$ are similar to those at 1.65 K. In Fig.~\ref{Fig-6}, we show the effective kinematic viscosity $\nu'$ obtained at different temperatures (blue triangles). To aid our discussions, we have also included in Fig.~\ref{Fig-6} the kinematic viscosity $\nu'$ calculated with our vortex density data using Eq.~(\ref{eqn-L}) (black squares), the $\nu'$ values obtained by Stalp \emph{et al.} in the towed-grid experiment \cite{Stalp-2002} (red solid circles), and the kinematic viscosity $\nu_n=\mu_n/\rho$ calculated based on the tabulated normal-fluid viscosity $\mu_n$ \cite{Donnelly-1998} (black solid curve). It is clear that $\nu_n$ is smaller than $\nu'$, which reflects the fact that the dissipation processes in quasiclassical turbulence in He II include not only the normal-fluid viscosity but also mutual friction. We note that the $\nu'$ values determined using our new methods appear to be greater than both the values calculated using the traditional method via Eq.~(\ref{eqn-L}) and those from Stalp \emph{et al.}. This difference may reflect the inherent limitations associated with the hypotheses involved in deriving the Eq.~(\ref{eqn-L}). Indeed, one can see clearly in Fig.~\ref{Fig-3} that the structure function at large decay times exhibits a peak at a scale smaller than the channel width, indicating the energy-containing eddy size being smaller than $D$. The black squares in Fig.~\ref{Fig-6} would appear much lower if a smaller energy-containing eddy size is used in the calculation. It is worthwhile noting that the $\nu'$ values have been determined in similar temperature range by Skrbek's group using Eq.~(\ref{eqn-L}) in the study of decaying counterflow \cite{Babuin-2015-PRB} and decaying bellow-induced co-flow turbulence \cite{Babuin-2014-JLTP}. Despite the large error bars, their data appear to be also greater than those from Stalp \emph{et al.} Nevertheless, without any information about the actual energy spectrum and energy-containing eddy size in these experiments, it is hard to comment on the reliability of these results.

\begin{acknowledgments}
We acknowledge the support from the US Department of Energy under Grant DE-FG02 96ER40952 and the National Science Foundation under Grant No. DMR-1507386. We would also like to thank D.N. McKinsey and S.W. Van Sciver for providing laser and cryogenics equipment.
\end{acknowledgments}


\begin{thebibliography}{99}
\bibitem{Tilley-1986-book} D.R. Tilley and J. Tilley, \emph{Superfluidity and Superconductivity} (2nd Ed, Adam Hilger, Bristol, 1986).
\bibitem{Donnelly-1998} R. J. Donnelly and C. F. Barenghi1, J. Phys. Chem. Ref. Data \textbf{27}, 1217 (1998).
\bibitem{Donnelly-1991-book} R.J. Donnelly, \emph{Quantized Vortices in Helium II}, (Cambridge University Press, Cambridge, England, 1991).
\bibitem{Vinen-2002-JLTP} W.F. Vinen and J.J. Niemela, J. Low Temp. Phys., \textbf{126}, 167-231 (2002).
\bibitem{Stalp-1999-PRL} S.R. Stalp, L. Skrbek, and R.J. Donnelly, Phys. Rev. Lett., \textbf{82}, 4831-4834 (1999).
\bibitem{Maurer-1998-EPL} J. Maurer and P. Tabeling, Europhys. Lett., \textbf{43}, 29 (1998).
\bibitem{Skrbek-2003-PRE} L.Skrbek, A.V. Gordeev, and F.Soukup, Phys. Rev. \textbf{E 67}, 047302 (2003).
\bibitem{Chagovets-2007-PRE} T.V. Chagovets, A.V. Gordeev, and L. Skrbek, Phys. Rev. \textbf{E 76}, 027301 (2007).
\bibitem{Gao-2015-JETP} J. Gao, W. Guo, V.S. L＊vov, A. Pomyalov, L. Skrbek, E. Varga, and W.F. Vinen, JETP Letters, \textbf{103}, 732 (2016).
\bibitem{Vinen-2000-PRB} W.F. Vinen, Phys. Rev. \textbf{B 61}, 1410 (2000).
\bibitem{Barenghi-1997-PhysFlu} C.F. Barenghi, D.C. Samuels, G.H. Bauer, and R.J. Donnelly, Phys. Fluids, \textbf{9}, 2361 (1997).
\bibitem{Baggaley-2012-EPL} A.W. Baggaley, C.F. Barenghi, A. Shukurov, and Y.A. Sergeev, Europhys. Lett., \textbf{98}, 26002 (2012).
\bibitem{Volovik-2003-JETP} G. E. Volovik, JETP Lett. \textbf{78}, 533 (2003).
\bibitem{Tsubota-2004-PRB} M. Tsubota, C.F. Barenghi, T. Araki, and A. Mitani, Phys. Rev. \textbf{B 69}, 134515 (2004).
\bibitem{Kivotides-2011-JFM} D. Kivotides, J. Fluid Mech., \textbf{668}, 58 (2011).
\bibitem{Walmsley-2014-PNAS} P. Walmsley, D. Zmeev, F. Pakpour, and A. Golov, Proc. Natl. Acad. Sci., \textbf{111}, 4691每4698 (2014).
\bibitem{Zmeev-2015-PRL} D.E. Zmeev, P.M. Walmsley, A.I. Golov, P.V.E. McClintock, S.N. Fisher, and W.F. Vinen, Phys. Rev. Lett., \textbf{115}, 155303 (2015).
\bibitem{Vinen-2001-PRB} W.F. Vinen, Phys. Rev. \textbf{B 64}, 134520 (2001).
\bibitem{Vinen-2003-PRL} W.F. Vinen, M. Tsubota, and A. Mitani, Phys. Rev. Lett., \textbf{91}, 135301 (2003).
\bibitem{Skrbek-1999-JP} L. Skrbek, J.J. Niemela, and R.J. Donnelly, J. Phys.: Condens. Matter, \textbf{11}, 7761-7781 (1999).
\bibitem{Donnelly-1991-Collection} R.J. Donnelly, \emph{High Reynolds Number Flows Using Liquid and Gaseous Helium}, (Springer-Verlag, New York, 1991).
\bibitem{Fuzier-2001} S. Fuzier, B. Baudouy, and S.W. Van Sciver, Cryogenics, \textbf{41}, 453-458 (2001).
\bibitem{Michel-2014} B. Saint-Michel, E. Herbert, J. Salort, C. Baudet, M.B. Mardion, P. Bonnay, M. Bourgoin, B. Castaing, L. Chevillard, F. Daviaud, P. Diribarne, B. Dubrulle, Y. Gagne, M. Gibert, A. Girard, B. Hebral, T. Lehner, and B. Rousset, Phys. Fluids, \textbf{26}, 125109 (2014).
\bibitem{Skrbek-2012-PhysFlu} L. Skrbek and K.R. Sreenivasan, Phys. Fluids, \textbf{24}, 055109 (2012).
\bibitem{Henze-1975-book} O. Henze, \emph{Turbulence}, 2nd ed. (McGraw-Hill, New York, 1975).
\bibitem{Stalp-2002} S. R. Stalp, J. J. Niemela, W. F. Vinen, and R. J. Donnelly, Phys. Fluids, \textbf{14}, 1377 (2002).
\bibitem{Niemela-2000-JLTP} J.J Niemela, K.R. Sreenivasan, and R.J. Donnelly, J. Low Temp. Phys., \textbf{138}, 537 (2000).
\bibitem{Babuin-2015-PRB} S. Babuin, E. Varga, W. F. Vinen, and L. Skrbek, Phys. Rev. \textbf{B 92}, 184503 (2015).
\bibitem{Walmsley-2008-PRL} P.M. Walmsley and A.I. Golov, Phys. Rev. Lett., \textbf{100}, 245301 (2008).
\bibitem{Walmsley-2007-PRL} P.M. Walmsley, A.I. Golov, H.E. Hall, A.A. Levchenko, and W.F. Vinen, Phys. Rev. Lett., \textbf{99}, 265302 (2007).
\bibitem{Bradley-2008-PRL} D.I. Bradley, S.N. Fisher, A.M. Guenault, R.P. Haley, S. O'Sullivan, G.R. Pickett, and V. Tsepelin, Phys. Rev. Lett., \textbf{101}, 065302, (2008).
\bibitem{Zmeev-2015-suppl} See discussions in the Supplemental Material in ref.~\cite{Zmeev-2015-PRL}.
\bibitem{Babuin-2014-EPL} S. Babuin, E. Varga, L. Skrbek, E. L谷v那que, and P. Roche, Europhys. Lett., \textbf{106}, 24006 (2014).
\bibitem{Vinen-1957-PRS} W.F.Vinen, Proc. Roy. Soc. \textbf{A 240}, 114 (1957); Proc. Roy. Soc. \textbf{A 240}, 128 (1957); Proc. Roy. Soc. \textbf{A 242}, 493 (1957); Proc. Roy. Soc. \textbf{A 243}, 400 (1958).
\bibitem{Awschalom-1984-PRL} D.D. Awschalom and K.W. Schwarz, Phys. Rev. Lett., \textbf{52}, 49 (1984).
\bibitem{Samuels-1998-PRL} D.C. Samuels and C.F. Barenghi, Phys. Rev. Lett., \textbf{81}, 4381 (1998).
\bibitem{Bradley-2011-Nat} D.I. Bradley, S.N. Fisher, A.M. Gu谷nault, R.P. Haley, G.R. Pickett, D. Potts, and V. Tsepelin, Nature Phys., \textbf{7}, 473 (2011).
\bibitem{Gao-2015-RSI} J. Gao, A. Marakov, W. Guo, B.T. Pawlowski, S.W. Van Sciver, G.G. Ihas, D.N. McKinsey, and W.F. Vinen, Rev. Sci. Instrum., \textbf{86}, 093904 (2015).
\bibitem{Marakov-2015-PRB} A. Marakov, J. Gao, W. Guo, S.W. Van Sciver, G.G. Ihas, D.N. McKinsey, and W.F. Vinen, Phys. Rev. \textbf{B 91}, 094503 (2015).
\bibitem{Landau-book} L.D. Landau and E.M. Lifshitz, \emph{Fluid Mechanics} (Pergamon Press, Oxford, UK, 1987).
\bibitem{Rellergert-2008-PRL} W.G. Rellergert, S.B. Cahn, A. Garvan, J.C. Hanson, W.H. Lippincott, J.A. Nikkel, and D.N. McKinsey, Phys. Rev. Lett., \textbf{100}, 025301 (2008).
\bibitem{Guo-2009-PRL} W. Guo, J.D. Wright, S.B. Cahn, J.A. Nikkel, and D.N. McKinsey, Phy. Rev. Lett., \textbf{102}, 235301/1-4 (2009).
\bibitem{Guo-2010-PRL} W. Guo, S.B. Cahn, J.A. Nikkel, W.F. Vinen and D.N. McKinsey, Phys. Rev. Lett., \textbf{105}, 045301 (2010).
\bibitem{Guo-2014-PNAS} W. Guo, M. La Mantia, D.P. Lathrop, and S.W. Van Sciver, Proc. Natl. Acad. Sci., \textbf{111}, 4653 (2014).
\bibitem{Gordeev-2005-JLTP} A.V.Gordeev, T.V.Chaguvets, F.Soukup, and L.Skrbek, J. Low Temp. Phys., \textbf{138}, 649 (2005).
\bibitem{Kolmogorov-1941} A.N. Kolmogorov, C. R. Acad. Sci. USSR, \textbf{32}, 16-18 (1941).
\bibitem{Frisch-1995-book} U. Frisch, \emph{Turbulence: The Legacy of A. N. Kolmogorov} (Cambridge Univ. Press, Cambridge, MA, U.S.A., 1995).
\bibitem{Schwarz-1991-PRL} K.W. Schwarz and J.R. Rozen, Phys. Rev. Lett., \textbf{66}, 1898 (1991).
\bibitem{Nemirovskii-1994-Cryo} S. Nemirovskii, L. Kondaurova, and M. Nedoboiko, Cryogenics, \textbf{34}, 309 (1994).
\bibitem{Sciacca-2010-PRB} M. Sciacca, Y. A. Sergeev, C. F. Barenghi, and L. Skrbek, Phys. Rev. \textbf{B 82}, 134531 (2010). 
\bibitem{Babuin-2014-JLTP} S. Babuin, E. Varga, and L. Skrbek, J. Low Temp. Phys., \textbf{175}, 324每330 (2014).
\end{thebibliography}
\end{document}